\documentclass[aps,prd,groupaddress,onecolumn,usenatbib,nofootinbib,showpacs,altaffilletter]{revtex4}
\usepackage{graphicx}% Include figure files
\usepackage{dcolumn}% Align table columns on decimal point
\usepackage{bm}% bold math
\usepackage{color}
\usepackage{amssymb,amsmath,latexsym,amsfonts}
\usepackage[none]{hyphenat}

\begin{document}

\title{Modified uncertainty principle from the free expansion of a Bose--Einstein Condensate}

\author{El\'ias Castellanos}
\email{ecastellanos@mctp.mx} 
\affiliation {Mesoamerican Centre for Theoretical Physics,\\ Universidad Aut\'onoma de Chiapas.
Ciudad Universitaria, Carretera Zapata Km. 4, Real del Bosque (Ter\'an), 29040, Tuxtla Guti\'errez, Chiapas, M\'exico.}

\author{Celia Escamilla-Rivera}
\email{cescamilla@mctp.mx}
\affiliation {Mesoamerican Centre for Theoretical Physics,\\ Universidad Aut\'onoma de Chiapas.
Ciudad Universitaria, Carretera Zapata Km. 4, Real del Bosque (Ter\'an), 29040, Tuxtla Guti\'errez, Chiapas, M\'exico.}

%--------------------------------------------------------------------------------------------------------------------------------------------
%--------------------------------------------------------------------------------------------------------------------------------------------
\begin{abstract}
In this paper we present a theoretical and numerical analysis 
of the free expansion of a Bose--Einstein condensate, where we assume 
that the single particle energy spectrum is deformed due to a possible quantum structure of space time. 
Also we consider the presence of inter particle interactions in order to study more realistic and specific scenarios. 
The \emph{modified} free velocity expansion of the condensate leads in a natural way to a modification of the 
uncertainty principle, which allows us to investigate some possible features of the Planck scale regime in low--energy earth--based experiments.\end{abstract}

%\date{}
\maketitle

%--------------------------------------------------------------------------------------------------------------------------------------------
%--------------------------------------------------------------------------------------------------------------------------------------------
\section{Introduction}
%\textit{Introduction.-}

Over the last few years, the use of many--body systems as theoretical tools in the search for possible Planck scale effects has become a very interesting line of research \cite{echa,Eli1,echa1,I,JD,AM}. In particular, systems related to Bose--Einstein condensates are promising pathways in the search for low-energy traces of Planck scale physics
\cite{echa,Eli1,echa1,CastellanosClaus,Castellanos,r1,r2,cam1,Conti,eli}.
These studies suggest that remarkable properties associated with Bose--Einstein condensates could be used to obtain representative bounds on the deformation parameters associated with quantum gravity models \cite{echa,Castellanos,r1,r2} or in specific cases
where it is possible to explore the sensitivity of these systems to Planck scale effects \cite{Eli1,echa1,CastellanosClaus,Conti,eli,poly}. 

An interesting phenomena related to Bose--Einstein condensates is the interference pattern when two condensates overlap \cite{Pethick,Andr,Mun}. The interference pattern is a manifestation of the wave (quantum) nature of these many--body systems and could be produced even when the two condensates are initially completely decoupled. After switching off the corresponding traps,
%, this allows 
the system expand whilst overlapping, eventually producing interference patterns \cite{Andr,JJ,YC,Mun}.

Furthermore, it is well know that when the trapping potential is turned off the free velocity expansion of the cloud corresponds to the velocity predicted by Heisenberg's uncertainty principle \cite{Pethick,Andr,Mun}. This fact is one of several reasons why Bose--Einstein condensates are relevant systems in the analysis and estimation of possible Planck scale effects, since quantum gravity models suggest modifications to this principle.\cite{KEN,AF,BM}.

Along these lines, in \cite{eli} we presented a study where the free velocity expansion of a Bose--Einstein condensate leads in a natural way to modifications of Heisenberg's uncertainty principle.  
If we assume as a fundamental fact that the energy per particle is modified due to the quantum structure of space time, then the predicted \emph{modified} free velocity expansion suggests a linear deformation in Heisenberg's uncertainty principle, 
\begin{equation}
 \Delta x \Delta p \ge  \frac{\hbar}{2} - \alpha^{*}x+ O(\alpha^{2}),
 \label{dup}
\end{equation}
where $\alpha^{*}=2\xi_{1}m^{2}c/3\sqrt{\pi} M_{p}$, $M_{p}$ is the Planck mass, $c$ is the speed of light, and $m$ is the mass of the particle. Additionally, $\xi_{1}$ is a real parameter of order unity, depending upon the quantum gravity model under consideration. 
%It is remarkable to notice that, 
As far as we know, this linear modification had not previously been reported in the literature, see for instance Refs.\,\cite{KEN,AF,BM}. 

The non--relativistic form of the aforementioned modified dispersion relation can be express in ordinary units as  follows \cite{Claus,Claus1}:
\begin{equation}
\epsilon \simeq
mc^2+\frac{p^{2}}{2m}+\frac{1}{2M_{p}}\Bigl(\xi_{1}mcp+\xi_{2}p^{2}+\xi_{3}\frac{p^{3}}{mc}\Bigr).
\label{ddr}
\end{equation}
The parameters $\xi_{1}$, $\xi_{2}$, and $\xi_{3}$, are model
dependent \cite{Giovanni1,Claus}, and should take\, positive or
negative values close to $1$ (see Ref.\,\cite{eli} for more details). In fact, the form of the energy dispersion relation (\ref{ddr}), was recently constrained by using high precision atom--recoil frequency measurements \cite{Claus,Claus1}. In this scenario, bounds for the deformation parameters of order $\xi_{1} \sim -1.8 \pm 2.1$ and 
$-3.8 \times 10^{9} < \xi_{2} < 1.5 \times 10^{9}$ were obtained.

Eq.\,(\ref{dup}) was deduced for a dilute system and neglecting the interactions among the particles within the condensate, i.e., the ideal case. Also, the \emph{modified} Heisenberg's uncertainty principle  is a consequence of  the leading order deformation contribution in Eq.\,(\ref{ddr}), which is linear in the momenta. 
In order to analyze the behaviour of the condensate under free expansion in the interacting case, more realistic scenarios are required. Clearly these must be taken into account when considering the corrections due by the next--to leading order term in Eq.\,(\ref{ddr}). 

With this aim, we will analyze the behaviour of the solutions to this \textit{modified} condensate scenario under free expansion using numerical tools, where we taken into account the effects produced by the leading order deformation, and the next--to leading order deformation in Eq. (\ref{ddr})  together with the interactions among the particles within the system extending the results reported in \cite{eli}.
%--------------------------------------------------------------------------------------------------------------------------------------------
\section{Free velocity expansion of the condensate}

The \emph{modified} energy associated with the system is given by
\begin{eqnarray}
\label{EN}
 E(\psi)&=&\int d \mathbf{r} \Bigg[\frac{\hbar^{2}}{2m}|\mathbf{\nabla} \psi(\mathbf{r})|^{2}+V(\mathbf{r})|\psi(\mathbf{r})|^{2} 
 +\frac{U_{0}}{2}|\psi(\mathbf{r})|^{4} + \hbar \alpha |\psi(\mathbf{r})\vert\sqrt{\vert\nabla\vert^{2}}\vert \psi(\mathbf{r})\vert+\beta \hbar^{2}|\mathbf{\nabla} \psi(\mathbf{r})|^{2}\Bigg],
\end{eqnarray}
where $\psi$ is the wave function of the condensate or the so--called order parameter, $V(r)=m\omega_{0}^{2}r^{2}/2$ is the external potential, which is assumed to be an isotropic harmonic oscillator for simplicity. The term $U_{0}=4\pi \hbar^{2}a/{m}$, depicts the interatomic potential, $a$ being the s--wave scattering length, i.e. only two--body interactions are taken into account. Notice that we
have included in the total energy of the cloud, the leading order modification in the deformed dispersion relation Eq.\,(\ref{ddr}), through the \emph{linear} operator $\vert\sqrt{\vert\nabla\vert^{2}}\vert$ \,\cite{eli,echa1}, where $\alpha=\xi_{1}mc/2M_{p}$. Also, we have inserted the next--to leading order deformation in Eq.\,(\ref{ddr}), through the usual operator $\mathbf{\nabla}$, corresponding to the deformation parameter $\beta=\xi_{2}/2M_{p}$. Notice that this term is also quadratic in the momenta as is the corresponding kinetic energy.
%Let us mention that 
The corrections caused by the deformation parameter $\beta$, could be \emph{re--absorbed} in the usual kinetic energy term by defining the effective mass $m_{\xi_{2}}=M_{p}m/(M_{p}+\xi_{2}m)$. We could also perform a similar analysis by assuming from the beginning that the deformation parameter $\beta$ is only a shift in the corresponding particle mass. However, as was pointed out in Ref.\,\cite{echa1}, both approaches lead to the same predictions for the ground--state energy and its properties, at least to first order in $\xi_{2}$. Thus, without loss of generality, we analyze in this work the modifications caused by $\beta$ as independent contributions to the total energy of the system.

 Furthermore, we have assumed that $\xi_{3}=0$. Some insights about this latter parameter in the case when is non--zero will be given at the end of this letter.
If we set $\alpha=\beta=0$ in the total energy Eq.\,(\ref{EN}) we recover the usual expression \cite{Pethick}.

The total energy of the cloud can be expressed as follows:
\begin{equation}
\label{TE}
E=E_{F}+E_{R},
\end{equation}
where $E_{F}$ is the kinetic energy associated with particle currents
\begin{equation}
\label{EFL}
E_{F}=\frac{\hbar^{2}}{2m} \int d\,\mathbf{r}|\psi(\mathbf{r})|^{2} (\mathbf{\nabla} \phi)^{2}.
\end{equation}

The function $E_{R}$ can be rewritten using the following components
\begin{equation}
E_{R}=E_{0}+E_{P}+E_{I}+E_{\alpha}+E_{\beta},
\end{equation}
where
\begin{eqnarray}
\label{EZP}
E_{0}&=&\frac{\hbar^{2}}{2m}\int d \mathbf{r} \Bigl(\frac{d |\psi (\mathbf{r})|}{dr}\Bigr)^{2}, \quad
E_{P}=\frac{1}{2}m\omega_{0}^{2}\int d \mathbf{r} r^{2} |\psi (\mathbf{r})| ^{2}, \nonumber
\end{eqnarray}
\begin{eqnarray} 
E_{I}&=&\frac{1}{2}U_{0}\int d \mathbf{r} |\psi (\mathbf{r})|^{4}, \quad
E_{\alpha}=\hbar \alpha \int d \mathbf{r} \Bigl(\frac{d |\psi (\mathbf{r})|^{2}}{dr}\Bigr),\nonumber \\
E_{\beta}&=&\beta \hbar^{2}\int d \mathbf{r} \Bigl(\frac{d |\psi (\mathbf{r})|}{dr}\Bigr)^{2},
\end{eqnarray}
so, $(E_{0})$ is related to the contributions of the ground state energy, ($E_{P}$) the contributions of the trapping potential,  and  $(E_{I})$ 
the contributions due to he particle interactions within the condensate. The contributions $(E_{\alpha})$ and $(E_{\beta})$ 
contains the contributions of the deformation parameters $\alpha$ and $\beta$, respectively.

Firstly, $E_{R}$ can be written as
\begin{eqnarray}
E_{R}&=&\frac{3}{4}\frac{{\hbar}^{2}}{m {R}^{2}}N+\frac{3}{4}m{{\omega}_{0}}^{2}{R}^{2}N
 +\frac{{U}_{0}}{2{(2\pi)}^{3/2} R^{3}}{N}^{2}
 - \alpha\frac{2 \hbar}{\sqrt{\pi}R}N +\beta\frac{3\hbar^{2}}{2R^{2}}N,
\end{eqnarray}
where we have used the ansatz
\begin{equation}
 \psi(\mathbf{r})=\frac{ N^{1/2}}{{\pi}^{3/4}R^{3/2}} \exp(-r^{2}/2R^{2})\exp[i\phi(r)], \label{TF}
\end{equation}
where $N$ is the corresponding number of particles
%\textbf{and as mentioned above
and $\phi(r)$ is a phase related to particle flows in the system.

The choice of the ansatz (\ref{TF}), for the case of a weakly interacting Bose--Einstein condensate trapped in an isotropic three--dimensional harmonic-oscillator potential, seems to be a good conjecture for several reasons: First, 
Eq. (\ref{TF}) clearly reflects the symmetry of the trap and in the non-interacting case is the exact solution of the corresponding equation of motion; Secondly, as was proven in the experiment described in \cite{Mun}, the system operates deeper in the linear regime for sufficiently large expansion times, i.e., the system evolves almost as in the non--interacting case in this situation. This fact further supports the use of the ansatz (\ref{TF}).
In other words, in the experiment \cite{Mun} was shown that the free velocity expansion at large times confirms that the evolution of the condensate can be independent of interactions during extended free fall experiments. Accordingly, the free velocity expansion can be computed in this scenario without loss of generality, by using the aforementioned ansatz at least to first order approximation in the deformation parameters $\alpha$ and $\beta$. Thirdly, as we will show later on this paper, all these facts indicate that large expansion times are also relevant in the search for some Planck scale effects.

%We note that the velocity and the phase are related through $\mathbf{\nabla} \phi= (m/\hbar)\mathbf{v}$ \cite{Pethick}, therefore the corresponding phase $\phi(r)$ must contain information of the deformation parameters due to the relation $v=d\,\epsilon/d\,p$ (see the deformed dispersion relation Eq.\,(\ref{ddr})). However, the order of magnitude of the corresponding correction in the velocity is at best of the order of $v\,(\alpha+\beta m v_{\beta})$, where $v_{\beta}=p/M_{p}$ as can be seen from Eq.\,(\ref{EFL}). 
Let us add that possible contributions due to the deformation terms can appear in the order parameter Eq.\,(\ref{TF}) through the corresponding phase $\phi(r)$, which is related to the \emph{local velocity of the condensate} as $\textbf{v}=(\hbar/m) \mathbf{\nabla} \phi(r)$ \cite{Pethick}. However, within our approach, all the measurable quantities of interest are computed by taking the norm of Eq.\,(\ref{TF}), which is related to the density and its derivatives (see Eqs.(\ref{EZP})). 
%Consequently, contributions of the deformation terms will only appears in the order parameter. 
%In general conditions, we are capable to analyze an equivalent set of equations for the density $|\psi|^{2}$, and the gradient of its phase, which is proportional to the \emph{local velocity} of the condensate 
 In other words, it is necessary to calculate the \emph{full solution} of the equation of motion, e.g., the corresponding Gross--Pitaevskii equation, together with the contributions due to the deformation terms.  This general version of Eq.\,(\ref{TF}) can be helpfully to analyze the contributions of the deformation terms upon the phase, the density and its derivatives. In order to test the validity of our model, the eventual predictions from the \emph{full solution} can be useful to compare with the approach presented in this work. This is a non--trivial topic that deserves deeper analysis and it will be presented elsewhere.

%In other words, within our approach all the representative properties which are related to measurable quantities, the possible modifications caused by the deformation parameters vanish within our approach, see Eqs.\,(\ref{EFL}) and (\ref{EZP}). 

Let us start with our model by considering that the external potential $V(r)$ is turned off at $t=0$, in such a case there is a force that changes $R$ and produces an expansion of the cloud \cite{Pethick}. It is straightforward to obtain the kinetic energy $E_{F}$ by using the \emph{ansatz} Eq.\,(\ref{TF}), with the result $E_{F}=3\dot{R}^{2}Nm/4$. Moreover, assuming that the energy is conserved at any time, we obtain the following energy conservation condition associated with our system
\begin{eqnarray}
\label{ERES}
\frac{3m \dot{R}^{2}}{4}+\frac{3\hbar^{2}}{4mR^{2}}+\frac{{U}_{0}N}{2{(2\pi)}^{3/2} R^{3}}
 -\alpha \frac{2 \hbar}{\sqrt{\pi}R}+ \beta\frac{3\hbar^{2}}{2R^{2}}=\frac{3\hbar^{2}}{4mR_{0}^{2}}+\frac{{U}_{0}N}{2{(2\pi)}^{3/2} R_{0}^{3}}
 -\alpha \frac{2 \hbar}{\sqrt{\pi}R_{0}}+ \beta\frac{3\hbar^{2}}{2R_{0}^{2}},
 \end{eqnarray}
where the dot stands for derivative with respect to time and $R_{0}$ is the radius of the condensate at time $t=0$, which is approximately equal to the oscillator
length $a_{ho}=(\hbar/m\omega_{0})^{1/2}$ in the non--interacting case. Otherwise, when interactions are present, we will assume that the initial radius corresponds to the result for an isotropic trap \cite{Pethick}
\begin{equation}
\label{RI}
R_{0}=\Bigl(\frac{2}{\pi}\Bigr)^{1/10}\Bigg(\frac{Na}{a_{ho}}\Bigg)^{1/5}a_{ho}.
\end{equation}

Additionally, $R$ is function of time and corresponds to the radius at time $t$. 
If we set $\alpha=\beta=0$ then we recover the usual solution in the non interacting case \cite{Pethick} which is given by 
\begin{equation}
\label{Usual}
{R}^{2}(t)={R}_{0}^{2}+\Bigl(\frac{\hbar }{m R_{0}}\Bigr)^{2}t^{2}.
\end{equation}

Notice that in the usual case, $\alpha=\beta=0$, $v_{0}=\hbar/mR_{0}$, is defined as the velocity expansion of the condensate, corresponding to the velocity predicted by Heisenberg's uncertainty principle for a particle confined within a distance $R_{0}$ \cite{Pethick}. Thus, in the usual case $\alpha=\beta=0$, the width of the cloud at time $t$ can be written in its usual form 
\begin{equation}
\label{usual}
{R}^{2}(t)={R}_{0}^{2}+(v_{0}t)^{2}.
\end{equation}

It is noteworthy to mention that when interactions are neglected we are able to obtain an analytical solution for Eq.\,(\ref{ERES}) when $R>>R_{0}$ together with $\alpha\ll1$ and $\beta\ll1$. In such s scenario we obtain
 \begin{eqnarray}
\label{SOLR2}
R_{\alpha,\beta}^{2}(t) = R_{0}^{2}+\Bigg[\frac{\hbar^{2}}{m^{2}R_{0}^{2}}\Bigl(1+2m\beta\Bigr)^{2}-\alpha \frac{8}{3 \sqrt{\pi} } \frac{\hbar}{m R_{0}}\Bigg]t^{2}.\,\,\,\,\,
\end{eqnarray}
If we set $\beta=0$, the result obtained in Ref.\cite{eli} is recovered. Thus, we may recognize the free velocity expansion in function of the deformation parameters $\alpha$ and $\beta$, which is given by
\begin{equation}
\label{VA}
(v_{0}^{\alpha,\beta})^{2}=\frac{\hbar^{2}}{m^{2}R_{0}^{2}}\Bigl(1+2 m\beta\Bigr)^{2}-\alpha \frac{8}{3 \sqrt{\pi} } \frac{\hbar}{m R_{0}}.
\end{equation}  
Since the corrections caused by $\alpha$ and $\beta$ are quite small the following expansion is justified:
\begin{equation}
(v_{0}^{\alpha,\beta})\approx \frac{\hbar}{mR_{0}}(1+2 m\beta)-\alpha \frac{4}{3 \sqrt{\pi}}.
\label{fve}
\end{equation}

Then, the velocity expansion corresponds to the following deformed Heisenberg's uncertainty principle
\begin{equation}
 \Delta x \Delta p \ge  \frac{\hbar}{2}(1+\beta^{*}) - \alpha^{*} x+\ldots,
 \label{dhp}
\end{equation}
where we have defined $\beta^{*}\equiv \xi_{2} m/ M_{p}$ and $\alpha^{*}\equiv \xi_{1}2m^{2}c/3\sqrt{\pi} M_{p}$ together with $R_{0}=x$.
It is not surprising that the functional form of  Eq.\,(\ref{dhp}) implies the following minimum measurable momentum and maximum measurable length 
\begin{equation}
\Delta p \ge (\Delta p_{min})\approx -\alpha^{*},
\label{minp}
\end{equation}
\begin{equation}
\Delta x \leq (\Delta x_{max}) \approx \frac{\hbar}{2}\Bigl(\frac{1+\beta^{*}}{\alpha^{*}}\Bigr).
\label{maxx}
\end{equation}
Notice that the inequality (\ref{minp}) is relevant only when $\xi_{1}<0$, which implies negative values of $\alpha^{*}$. 
These conditions also set the value range of deformation parameters $\xi_1$
and $\xi_2$ without breaking the inequality (\ref{dhp}). From a phenomenological point of view, these conditions can be used in other systems, in order to explore some issues related with the quantum structure of space time.

%--------------------------------------------------------------------------------------------------------------------------------------------
%--------------------------------------------------------------------------------------------------------------------------------------------

\section{Numerical Analysis}
%\textit{Numerical Analysis.- }

In order to explore the velocity expansion and the possible corrections caused by the deformation parameters 
$\xi_1$ and $\xi_2$, we need to solve Eq.\,(\ref{ERES}) numerically at any time and taking into 
account the interactions among the particles within the system. We will take fiducial laboratory conditions over the parameters related 
to the model as, $N\sim 10^{4}-10^{6}$ particles, $\omega_{0} \sim 10-10^{6}$ Hz, $a \sim10^{-9}$ m, and $m\sim 10^{-26}$ kg \cite{Dalfovo}. Additionally, $M_{p} \simeq 2.18 \times10^{-8} $kg, $\hbar\sim 6.623 \times 10^{-34}$Jás, $c\sim 3\times 10^{8}$m/s. The deformation parameters considered here will be of the order of  $\xi_{1} \sim -1.8 \pm 2.1$ and 
$-3.8 \times 10^{9} < \xi_{2} < 1.5 \times 10^{9}$. 

\begin{figure}[htbp]
\centering
\includegraphics[width=9.cm]{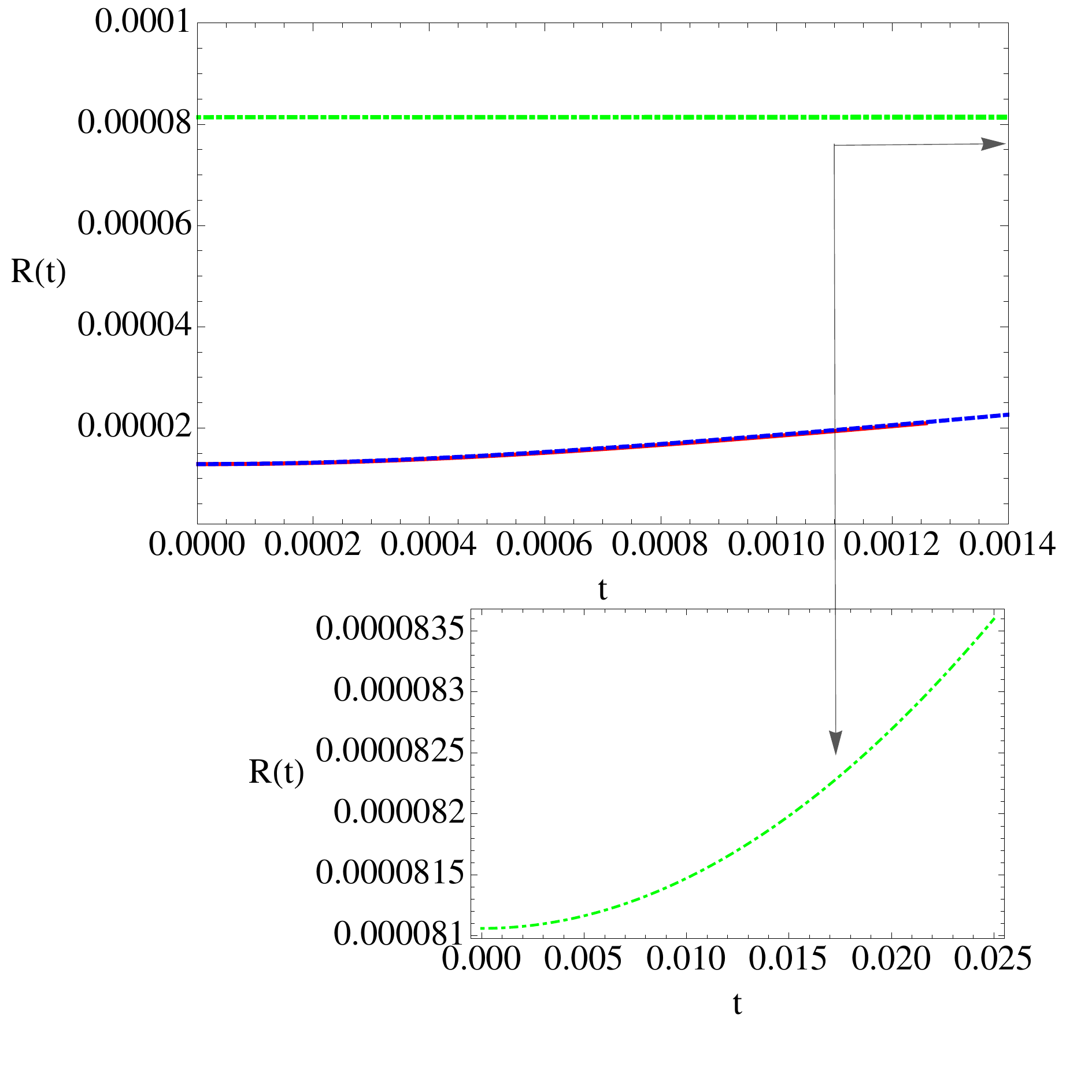}
\caption{\label{fig:General_R} Case [I]: For $\alpha, \beta$ and $U_0$ non--zero is represented by the red solid line. Case [II]: For $\alpha=\beta =0$ and $U_0=4\pi h^2 a/m$ is represented by the blue dashed line. Case [III]: For $\alpha, \beta, U_0 =0$ is represented by the green dashed--dotted line. The plot below shows the late time solution for Case [III].}
\end{figure}

To illustrate how all these ingredients work together properly, we study now the accuracy of our numerical
solution in some particular cases. These cases are: [I] For $\alpha, \beta$ and $U_0$ non-zero. [II] For
$\alpha=\beta =0$ and $U_0=4\pi h^2 a/m$ and [III] For $\alpha, \beta, U_0 =0$. Also, we impose initial conditions
for these cases. The condition at $t=0$ for the cases [I] and [II] is described by the Eq.(\ref{RI}). The condition 
for the Case [III] is set by $R(t=0)=(h/mw_0)^{1/2}$.
In Figure \ref{fig:General_R} we show the numerical solutions for $R(t)$. Cases [I] and [II], (red and blue lines, respectively) show an
identical evolution. The fact that the numerical solution for the Case [I] looks cut, shows that the code remains 
convergent a large times. Case (III) represented by the green dashed-dotted line gives the exact solution $R(t)=\sqrt{6.571\times 10^{-9} + 6.676 \times 10^{-7} t^2}$. The solution for this latter is illustrated in
the the plot inside Figure 1. 

%\begin{widetext}
\begin{figure*}
%\minipage{0.49\textwidth}
\centering
\includegraphics[width=10.0cm]{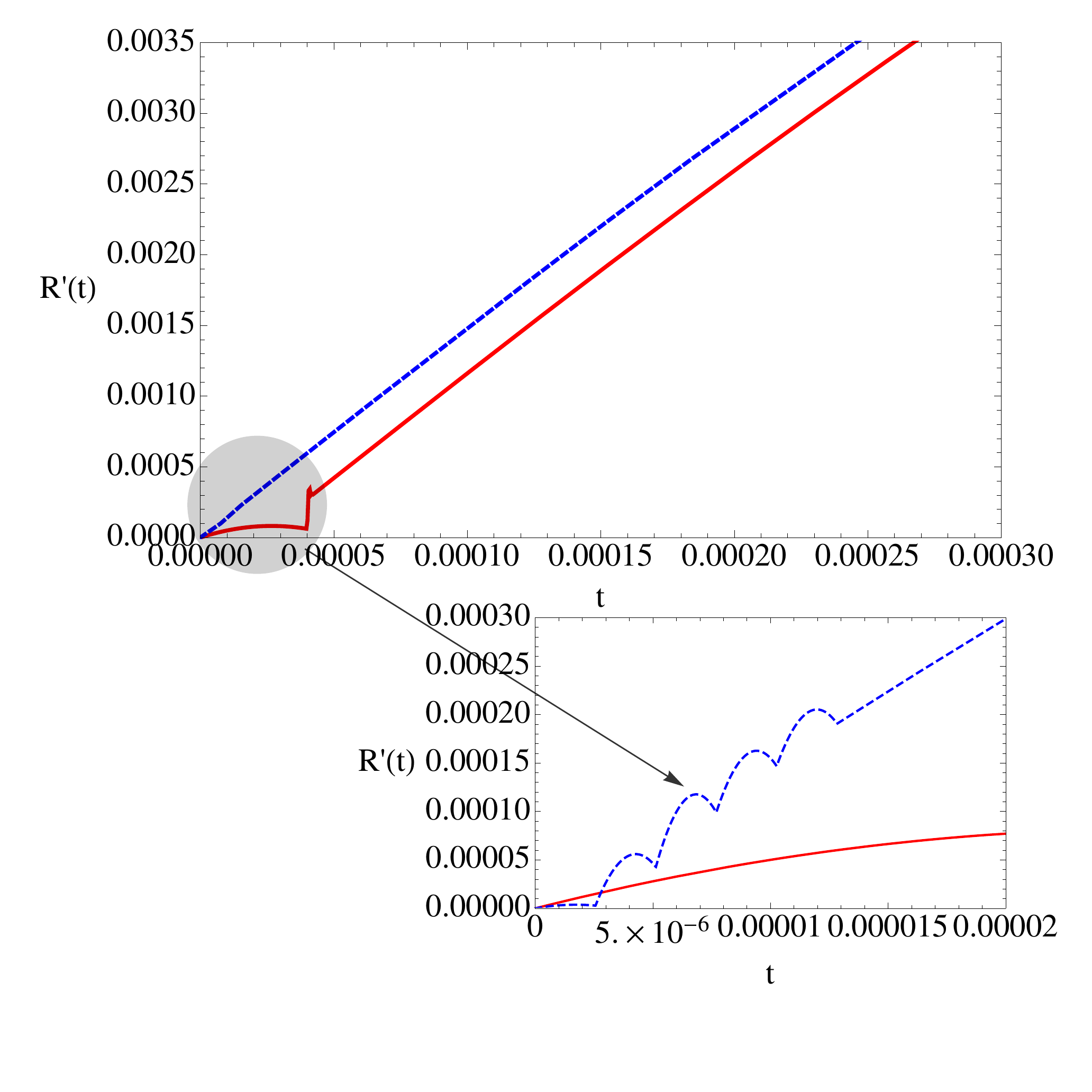} 
%(8.6, GeneralPlotF,GeneralPlotC)
\includegraphics[width=7.8cm]{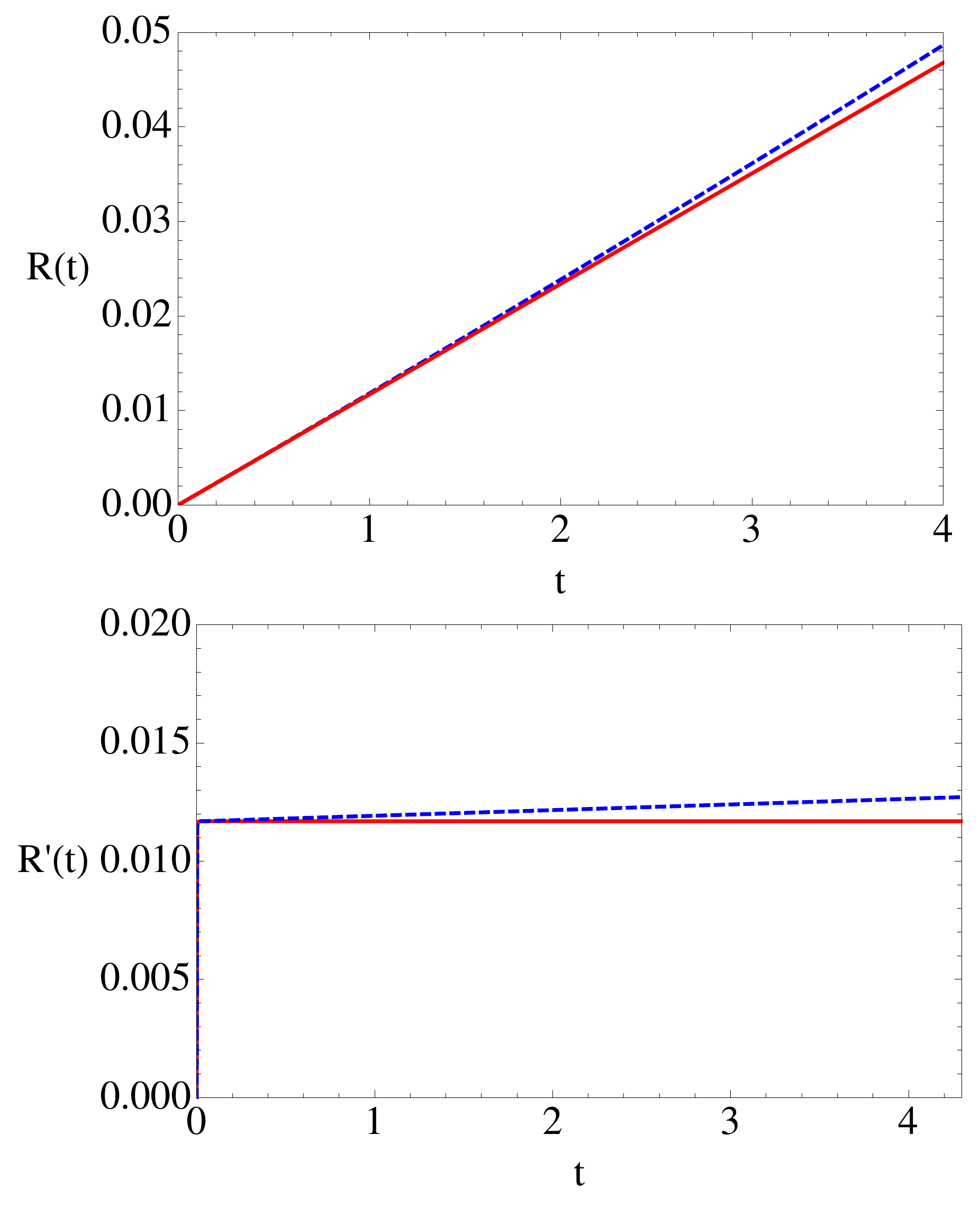} 
\centering
\caption{\label{fig:General_Rp1} \textit{Left}: For $R^\prime(t)$ at early times only the cases [I] and [II] are shown due the exact solution for Case [III]. Being so, Case [I] correspond to the red solid line and Case [II] correspond to the blue dashed line. 
The plot below represents the evolution at early times. \textit{Right}: For $R^\prime(t)$ at late times $t>4$ sec, the plot below represents the evolution of $R(t)$ in the same time range.}
%\endminipage
\end{figure*}
%\end{widetext}

In Figure \ref{fig:General_Rp1} are illustrated the numerical solutions for the modified velocity $R^\prime(t)$ for the cases [I] and [II]. The velocity in Case [III] is linear in time and its solution is exact, which for our interest in this figure with only show the modified cases.

We notice interesting points related to the modified velocity and the computation of \textit{modified} Heisenberg's uncertainty principle in each scenario:
\begin{itemize}
\item At early times, Eq.\,(\ref{dhp}) shows a value around $\Delta x \Delta p \gtrsim1.96 \times 10^{-36}$. This range correspond to the
point where the numerical solutions of $R'(t)$ for the cases [I] and [II] overlap as we see from the left plot inside Figure \ref{fig:General_Rp1}. This overlapping can be due to the equal dominance of both deformed parameters $\xi_1$ and $\xi_2$.

\item At the left of Figure.\,\ref{fig:General_Rp1} we observed a bounce in the solutions at early times. During this stage we have
for Case [I]: $\Delta x \Delta p \gtrsim 1.28 \times 10^{-35}$ and for Case [II]: $\Delta x \Delta p \gtrsim 4.38 \times 10^{-36}$.

\item After this instability, the evolution of the \textit{modified} velocity shows for Case [I]: $\Delta x \Delta p \gtrsim 4.59 \times 10^{-34}$ and for Case [II]:  $\Delta x \Delta p \gtrsim 7.22 \times 10^{-36}$. We observed a linear evolution of the \textit{modified} velocity in where the \textit{modified} Heisenberg's uncertainty principle for Case [I] is larger than Case [II] due the eventual dominance of the deformed parameters $\xi_1$ and $\xi_2$. This results was expected due the appearance of these deformations at small scales.

\item At late times, let us consider, e.g. $t\approx 4$ sec, as in free fall experiments \cite{Mun}. 
In this scenario we obtain for large times that $\Delta x \Delta p \gtrsim 10^{-30}$ (see at the right of Figure.\,\ref{fig:General_Rp1}). The evolution of the \textit{modified} velocity for Case [I] is a constant. This behaviour is in agreement with the theoretical result that at large times the corrections caused by the deformation parameters could be representative and consequently,  can be described by the \emph{modified} free velocity Eq.\,(\ref{fve}) when interactions are neglected \cite{eli}.
\end{itemize}

Generally, we notice that at very early times of expansion there is a period in which the velocity seems to be dominated by the deformation parameters. However, we estimate that this short period of expansion of order $t=7.744\times 10^{-6}$ sec may be hardly accessible from the experimental point of view, since some results offer order of milliseconds \cite{Andr}, i.e., three orders of magnitude bigger than the expansion time obtained here. Conversely, for large expansion times up to $t=4$ sec, there is a region in which the presence of the deformation parameters modified the velocity expansion in a way that may be significant, even when interactions are present.

Concerning the experiment performed in \cite{Mun}, it was proven that for sufficiently large expansion times, the system operates deeper in the linear regime, i.e.,  in the non-interacting case. In consequence,  the observed spatial interference pattern indicates that the fringe spacing scales linearly with the time of expansion and is inversely proportional to the initial separation of two condensates. In this experiment it was shown that the free velocity expansion at large times, confirms that the evolution of the condensate can be independent of interactions during extended free fall experiments.
Each of the above scenarios shows that the \emph{modified} free velocity expansion leads to deformations of Heisenberg's uncertainty principle which are around two orders of magnitude smaller than the typical case. This fact, could be tested, in principle, in the laboratory, if the possible corrections in the free velocity expansion of the condensate can be eventually be measured. However, let us remark that according to our results, large expansion times are required. This analysis opens a very important branch of research concerning the search of some quantum gravity traces in low energy earth based experiments.    

%--------------------------------------------------------------------------------------------------------------------------------------------
%--------------------------------------------------------------------------------------------------------------------------------------------
\section{Conclusions}
We have analyzed and described the free velocity expansion of a Bose--Einstein condensate at different times and also when interactions are present assuming a deformed dispersion relation as a fundamental fact. Additionally, we have obtained a deformation of Heisenberg's uncertainty principle which appears naturally just by looking up the \emph{modified} free velocity expansion.
However, for a further insight into this deformation the third deformation term in Eq.\,(\ref{ddr}) must also be taken into account. 
According to \cite{Landau} this cubic term could be interpreted as inversely proportional to the \textit{lifetime} of the condensate if we assume that this contribution to the total energy is imaginary. These facts, lead us to think that some of the particles leave the system (the condensate). In other words, this last assertion suggests that some of the particles forming the condensate, may be transferred to the excited states and in consequence could lead to instabilities within the system at some given time. 
Moreover, this deformation would contribute also to the functional form of the deformed uncertainty principle. These are non-trivial topics which deserve  deeper investigation and on which we will report elsewhere. 

Finally, according to our results there are two relevant scales of time associated with the free expansion, which 
offers a possibility to detect small signals or traces from the quantum structure of space time. However, we stress that an optimal scenario in searching these possible signals is when the system expands for large times. As we mentioned, free fall experiments 
could provide signs of Planck scale physics in this scenario.

%--------------------------------------------------------------------------------------------------------------------------------------------
%--------------------------------------------------------------------------------------------------------------------------------------------
\begin{acknowledgments}
%\textit{Acknowledgments.-}
E. C.  acknowledges MCTP for financial support and C. E-R. thanks CNPq Fellowship for support. We thanks to P. Sloane for
his opinion on the manuscript.
\end{acknowledgments}

%--------------------------------------------------------------------------------------------------------------------------------------------
%--------------------------------------------------------------------------------------------------------------------------------------------

\end{document}